\documentclass[twocolumn]{article}

\usepackage{graphicx}
\usepackage{float}
\usepackage{url}
\usepackage{authblk}
\usepackage{gensymb}

\title{\textbf{Backgrounds for Fast Simulation $\mathbf{e^+ e^-}$ Collider Studies at $\mathbf{\sqrt{s}=91,250,350,500}$ GeV}}
\author{C.T. Potter}
\affil{University of Oregon and Alder IHEP}
\date{\today}

\begin{document}

\maketitle

\begin{@twocolumnfalse}
\abstract{Various proposals for a new $e^+ e^-$ collider operating below 1 TeV are now under consideration by the worlwide High Energy Physics community. Among these are the International Linear Collider and the Circular Electron Positron Collider. We describe high statistics Standard Model background simulation samples generated with MG5\_aMC@NLO for $e^+e^-$ collider studies at $\sqrt{s}=91,250,350,500$~GeV. Fast detector simulation is performed with Delphes and DSiD, the detector card for the SiD detector. The samples are compared with other simulation samples generated with Whizard.}
\end{@twocolumnfalse}

\section{Introduction}

With the discovery of the Standard Model (SM) Higgs boson at the Large Hadron Collider (LHC) \cite{Aad:2012tfa,Chatrchyan:2012ufa}, the case for a new $e^+ e^-$ collider for precision Higgs measurements is strong \cite{Baer:2013cma,Dawson:2013bba,Asner:2013psa}. The clean $e^+e^-$ environment, where the initial state is well known and can be used to enhance final state measurements, contrasts with the more challenging environment at the LHC. The case is made even stronger by the potential for top quark measurements, which until now have only been made at hadron colliders. Possible new physics below 1 TeV makes the $e^+e^-$ collider case even stronger.

Two proposals, with somewhat complementary energy regimes, have been made for an $e^+e^-$ collider sited in Asia. The International Linear Collider (ILC) baseline design \cite{Adolphsen:2013kya} calls for a linear machine with $\sqrt{s}=500$~GeV sited in Japan.  The Circular Electron Positron Collider (CEPC) baseline design \cite{CEPC-SPPCStudyGroup:2015esa} calls for a circular machine with $\sqrt{s}=250$~GeV sited in China. 

Full simulation studies with comprehensive $e^+e^-$ collider backgrounds have been made for SM Higgs boson, top quark and new physics scenarios. But the background samples are in many cases statistically limited by the computing demands of full detector simulation, motivating fast detector simulation. The Delphes fast detector simulator \cite{deFavereau:2013fsa,Selvaggi:2014mya,Mertens:2015kba,Cacciari:2011ma}, which has been used extensively for LHC studies, is a generic detector simulation suitable for use in $e^+e^-$ studies which uses \texttt{tcl} text files to describe a particular detector's performance. 

In this study we use MG5\_aMC@NLO 2.3.3 \cite{Alwall:2014hca} to generate $e^+e^-,e\gamma$ and $\gamma\gamma$ processes for $\sqrt{s}=91,250,350$ and $500$~GeV. At each $\sqrt{s}$ separate backgrounds are produced for two beam polarization configurations, $e_{R}^{-}e_{L}^{+}$ and $e_{L}^{-}e_{R}^{+}$. At $\sqrt{s}=91,250$~GeV additional backgrounds are produced with unpolarized beams.  For the polarized beam samples, the integrated luminosities are approximately five times the projected target luminosities of the current ILC run scenarios. For the unpolarized beam samples at $\sqrt{s}=91$~GeV, the integrated luminosity is five times a GigaZ production, while for $\sqrt{s}=250$~GeV the integrated luminosity matches the statistics of the polarized beam samples. For validation, the polarized MG5\_aMC@NLO samples are compared with the samples generated with Whizard \cite{Kilian:2007gr} for the ILC Detailed Baseline Design (DBD) study \cite{Behnke:2013lya}. 

Fast detector simulation is performed on both the MG5\_aMC@NLO samples and DBD Whizard samples using  Delphes and the DSiD detector card \cite{Potter:2016pgp}. DSiD is modeled on the full simulation performance of the SiD detector as described in \cite{Behnke:2013lya} and is available on HepForge \cite{Buckley:2006nm} at \texttt{dsid.hepforge.org}. SiD was designed as a detector for the ILC, but it has also recently been proposed as a detector for CEPC \cite{Chekanov:2016efe}.

\begin{table*}[t!]
\begin{center}
\begin{tabular}{|c|c|c|c|} \hline
$\sqrt{s}$ [GeV] & G-20 ($+,-$)/($-,+$) & H-20 ($+,-$)/($-,+$) & I-20 ($+,-$)/($-,+$) \\ \hline
250  & 0.339/0.113 & \textbf{1.350/0.450} & 0.339/0.113 \\
350  & 0.135/0.045 & 0.135/0.045 & \textbf{1.146/0.382} \\
500  & \textbf{2.000/2.000} & 1.600/1.600 & 1.600/1.600\\ \hline
\end{tabular}
\caption{ILC integrated luminosity sharing (in ab$^{-1}$) for scenarios G-20, H-20 and I-20 defined in \cite{Barklow:2015tja}. For each $\sqrt{s}$, the integrated luminosities for the most optimistic scenario are in boldface.} 
\label{tab:scenarios}
\end{center}
\end{table*}

\section{Operating Scenarios}

The run program for an $e^+ e^-$ collider must be optimized for the set of measurements which various center-of-mass energies, beam polarizations, and integrated luminosities can provide. This can only be done when the physics landscape below 1~TeV is illuminated by the LHC, but assuming that mostly SM particles will be produced at lower energies and any possible BSM particles will be produced at higher energies, a first optimization can be made.

Precision $Z$ boson studies motivate running at the $Z$ pole. Precision SM Higgs boson measurements motivate running scenarios for $\sqrt{s}=250$~GeV, near the maximum cross section for $e^+e^- \rightarrow Zh_{SM}$ production, while precision top quark measurements motivate running scenarios for $\sqrt{s}=350$~GeV, just above threshold for $e^+e^- \rightarrow t\bar{t}$. In both cases background from $e^+ e^- \rightarrow W^+ W^-$ can be greatly reduced by colliding righthanded electrons with lefthanded positrons ($e_{R}^{-}e_{L}^{+}$) since $t$-channel production only occurs in interactions of lefthanded electrons with righthanded positrons ($e_{L}^{-}e_{R}^{+}$). At higher center-of-mass energies like $\sqrt{s}=500$~GeV, BSM physics motivates a more democratic luminosity sharing between beam polarizations. For  many BSM scenario signals background from $e^+ e^- \rightarrow W^+ W^-$ is not problematic, and the precision measurement of the chiral structure of new couplings argues for equal luminosity sharing of beam polarization configurations.

The physics goals for the CEPC, precision Higgs and $Z$ boson studies, motivate lower energies ($\sqrt{s}=91,250$~GeV). For the ILC, precision top quark and new physics studies motivate higher energies. The report issued by the ILC Parameters Joint Working Group \cite{Barklow:2015tja} identifies three operating scenarios based on these and other considerations: G-20, H-20 and I-20. All three scenarios envision a 20 year lifetime of the collider with one luminosity upgrade. 

G-20, motivated more by new physics than SM measurements, envisions most datataking at higher $\sqrt{s}$. H-20 and I-20 envision considerably more dataking at lower $\sqrt{s}$ than G-20. In the H-20 scenario most lower energy datataking occurs at $\sqrt{s}=250$~GeV, enhancing the precision of SM Higgs measurements, while in I-20 most low energy datataking occurs at $\sqrt{s}=350$~GeV. In G-20 the beam polarization configuration is democratic, while in H-20 and I-20 the $e_{R}^{-}e_{L}^{+}$ configuration is preferred over $e_{R}^{-}e_{L}^{+}$ by a ratio 3:1. For the ILC at $\sqrt{s}=250,350$~GeV, 10\% of luminosity is reserved in all three scenarios for $e_{L}^{-}e_{L}^{+}$ and $e_{R}^{-}e_{R}^{+}$ beam configurations, while 20\% is reserved for these configurations at $\sqrt{s}=500$~GeV. 

The proposed luminosity sharing (in ab$^{-1}$) between beam configurations $e_{L}^{-}e_{R}^{+}$ and $e_{R}^{-}e_{L}^{+}$ for the three scenarios is shown in Table~\ref{tab:scenarios}. In Table~\ref{tab:scenarios} and hereafter, $e_{R}^{-}e_{L}^{+}$ is denoted $(-,+)$ while $e_{L}^{-}e_{R}^{+}$ is denoted $(+,-)$\footnote{This notation differs from \cite{Barklow:2015tja}, where the $e^-$ handedness precedes the $e^+$ handedness, \emph{ie} $(+,-) \rightarrow (-,+)$ and \emph{vice versa}.}. Throughout this note, the ILC nominal 80\% electron and 30\% positron polarization is assumed.

\section{Standard Model Processes}

\begin{figure*}[t!]
\begin{center}
\begin{minipage}{0.45\textwidth}
\vspace{3.7in}
\includegraphics[height=\textwidth,angle=-90]{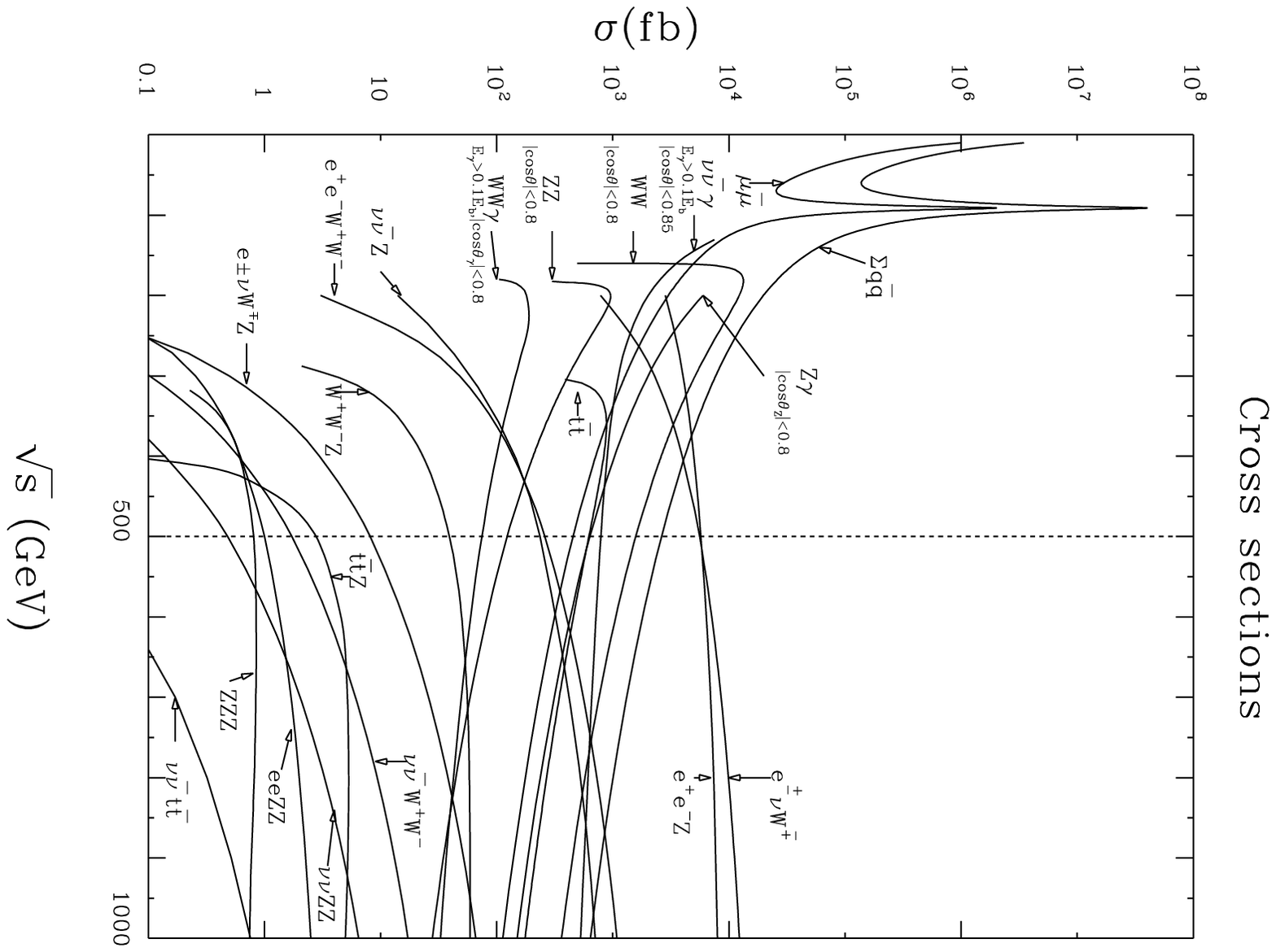}
\end{minipage}
\begin{minipage}{0.45\textwidth}
\includegraphics[width=\textwidth]{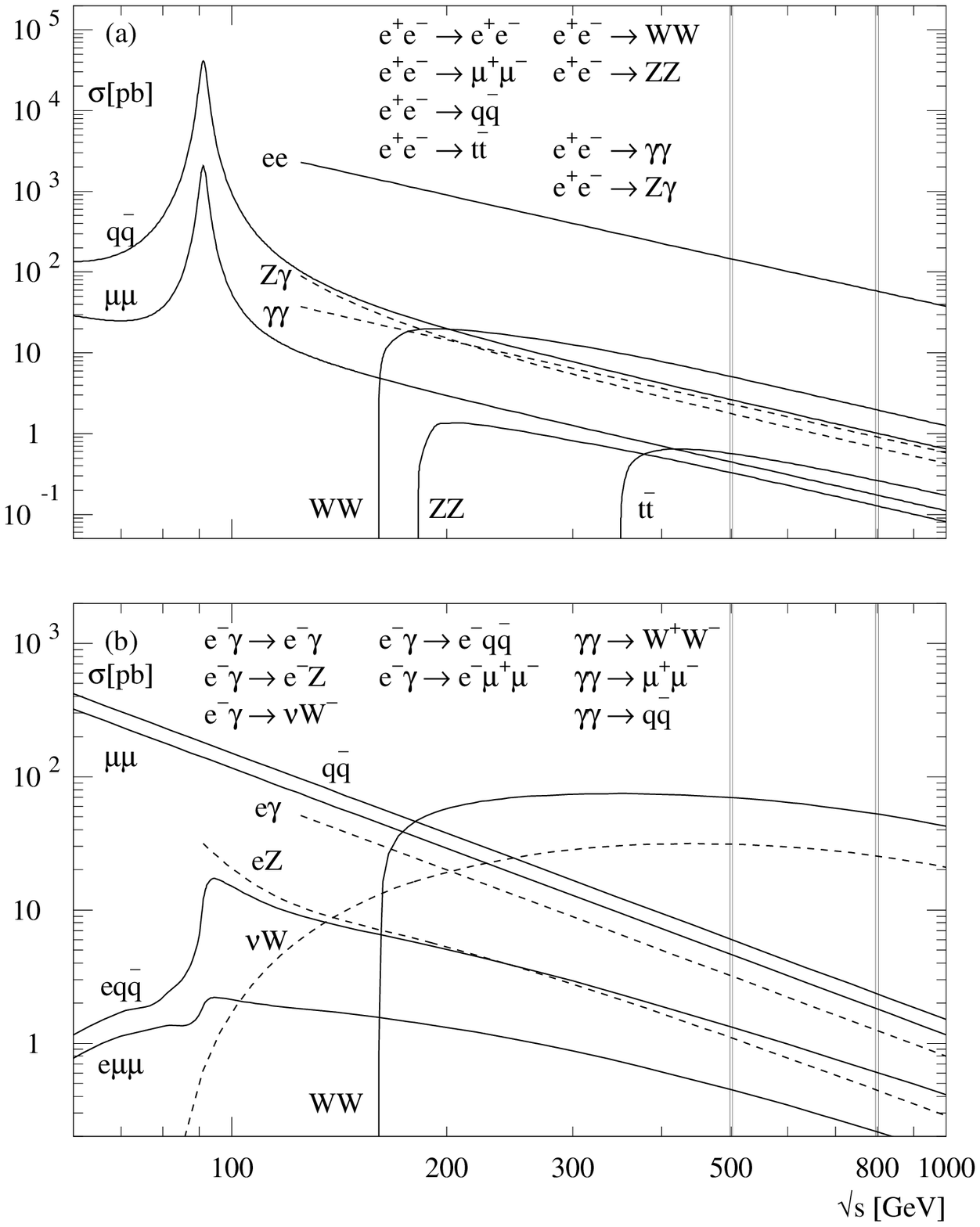}
\end{minipage}
\caption{Cross section vs $\sqrt{s}$ for unpolarized $e^+e^-,e\gamma$ and $\gamma \gamma$ initial states from \cite{Murayama:1996ec} (left) and \cite{Accomando:1997wt} (right). For the $e\gamma$ and $\gamma \gamma$ processes, the $\sqrt{s}$ refers to the $e\gamma$ and $\gamma \gamma$ center of mass energy.}
\label{fig:xs}
\end{center}
\end{figure*}

SM backgrounds and their cross section at $e^+e^-$ colliders have been discussed in \cite{Murayama:1996ec,Accomando:1997wt} and elsewhere. SM background processes for any $e^+e^-$ collider can be classified by center of mass energy, beam polarization and initial and final states. 

\begin{table}[h!]
\begin{center}
\begin{tabular}{|l|c|} \hline
Type & Process \\ \hline
2f,4f,6f & $e^+e^- \rightarrow 2f,4f,6f$ \\
1f,3f,5f & $e \gamma \rightarrow e \gamma,e2f,\nu2f,e4f,\nu4f$  \\
aa 2f,4f& $\gamma \gamma \rightarrow 2f,4f$  \\ \hline
\end{tabular}
\caption{Background typology for $e^+e^-$ colliders.} 
\label{tab:types}
\end{center}
\end{table}

Inital states include $e^+e^-,e\gamma$ and $\gamma \gamma$, where the $\gamma$ in an initial state originates from bremstrahlung. The final states are categorized by the number of fermions $f$ (1$f$, 2$f$, 3$f$, 4$f$, etc.) after boson decay. See Table~\ref{tab:types} for SM background typology and Figure~\ref{fig:xs} for the SM background process cross section as a function of $\sqrt{s}$ for unpolarized beams.

The SM background simulation for the CEPC run scenario have been described in detail \cite{Mo:2015mza}, where event generation is performed with Whizard with detailed ISR and bremstrahlung simulated with GuineaPig \cite{Schulte:1997nga}.  For the ILC run scenarios, SM background simulation has been described in detail in \cite{Behnke:2013lya}.  For full simulation benchmark studies for DBD study, SM background samples with integrated luminosities of 250fb$^{-1}$ were generated for each $\sqrt{s}=250$, $350$, $500$~GeV with Whizard 1.40 using Pythia6 \cite{Sjostrand:2006za} for showering and hadronization and saved in StdHEP format \cite{stdhep}. The samples were generated with 100\% lefthanded or righthanded electrons and positrons, from which new mixed samples were made assuming 30\% positron beam polarization and 80\%  electron beam polarization.

All SM background processes are included in the DBD samples. Beam conditions and backgrounds specific to ILC design parameters and bunch structure were generated with GuineaPig and passed to Whizard, including detailed beamstrahlung and bremstrahlung processes with the resulting beam energy distribution. In addition to the processes in Table \ref{tab:types}, the DBD samples also include pileup from bunch-bunch interactions: $\gamma \gamma$ to hadronic mini-jets and other low $p_T$ hadrons, as well as low $p_T$ beam-induced $e^+e^-$ pairs.
Both the CEPC and ILC DBD samples include interference effects. Whizard includes all diagrams producing the same final states through distinct intermediate particles and therefore includes the interference between these diagrams. For example, some final states $f \bar{f} f^{\prime} \bar{f^{\prime}}$ can be produced either by $ZZ$ or $W^+ W^-$ intermediate states. When specifying the fermion final states Whizard includes their interference.
In order to avoid divergent cross sections, kinematic cuts are imposed on the DBD Whizard samples during generation. The invariant mass of a pair of colored particles is required to be at least 10~GeV, while for a pair of colorless particles it is required to be at least 4~GeV. The minimum $\sqrt{-q^2}$ for q massless t-channel process is required to be 4~GeV.

\begin{table*}[t]
\begin{center}
\begin{tabular}{|l|c|c|c|c|c|} \hline
DBD Sample & $\sqrt{s}$ [GeV] & Pol. & $N$[M] & $\langle W \rangle$ & $\int  \mathcal{L}$ [ab$^{-1}$] \\ \hline \hline
\texttt{higgs\_ffh}  & 250 & ($+,-$) & 0.3 & 0.294 & 0.250 \\ 
\texttt{higgs\_ffh}  & 250 & ($-,+$) &  0.3 & 0.190 & 0.250 \\ 
\texttt{all\_SM\_background} & 250 & ($+,-$) &  2.8 & 255.9  &  0.250 \\ 
\texttt{all\_SM\_background} & 250 & ($-,+$) &  2.1 & 342.4 &  0.250 \\ \hline
\texttt{ttbar} & 350 & ($+,-$) & 0.3 & 1 & 1.000 \\
\texttt{ttbar} & 350 & ($-,+$) & 0.1 & 1 & 1.000 \\
\texttt{all\_other\_SM\_background}& 350 & ($+,-$) & 4.0 & 230.5 & 0.250 \\ 
\texttt{all\_other\_SM\_background}& 350 & ($-,+$) & 3.1 & 294.1 & 0.250 \\ \hline
\texttt{6f\_ttbar\_mt173p5} & 500 & ($+,-$) & 0.9 & 0.291  & 0.250 \\
\texttt{6f\_ttbar\_mt173p5} & 500 & ($-,+$) & 0.4 & 0.286  & 0.250 \\
\texttt{all\_SM\_background} & 500 & ($+,-$) & 2.3 & 536.8 & 0.250 \\ 
\texttt{all\_SM\_background} & 500 & ($-,+$) & 1.6  & 761.1 & 0.250 \\ \hline
\end{tabular}
\caption{The DBD samples \cite{Behnke:2013lya} used for comparison to samples in this study. Mean event weights $\langle W \rangle$ of order $10^2$ are due mostly to $4f$ processes with weight 12.5, $3f$ processes with weight 125 and the $1f$ process with weight 12,500.} 
\label{tab:dbd}
\end{center}
\end{table*}

Events in the DBD samples are weighted. Since some processes have prohibitively large cross sections, these events have large weights $W$ in order to reach the target integrated luminosity. Some signal processes of interest are weighted with small weights in order to provide a high statistics sample for study. Mean event weights $\langle W \rangle$ of order $10^2$ are due mostly to $4f$ processes with weight 12.5, $3f$ processes with weight 125 and the $1f$ process with weight 12,500. See Table \ref{tab:dbd} for the DBD samples used to compare with the backgrounds described in this note.\footnote{For more details and the DBD Whizard StdHEP files and logfiles see \url{https://confluence.slac.stanford.edu/} \url{display/ilc/Standard+Model+Data+Samples}.}

\section{Generation and Simulation}

The background samples in this study were generated with MG5\_aMC@NLO 2.3.3 \cite{Alwall:2014hca} with showering and hadronization by Pythia6 libraries implemented in the pythia-pgs package. While MG5\_aMC@NLO can calculate higher order corrections, only leading order samples were generated. Polarized samples for $\sqrt{s}=91,250,350$ and $500$~GeV are generated with electron polarization fixed to $\pm$80\% and positron polarization fixed to $\mp$30\%. Additional samples at $\sqrt{s}=91,250$~GeV are generated with unpolarized beams. For a summary of the background MG5\_aMC@NLO samples see Tables~\ref{tab:250},\ref{tab:91},\ref{tab:350},\ref{tab:500}.

Rather than specifiying multiple fermion final states, as was done for the DBD samples, intermediate top pair, diboson and triboson states are specified in MG5\_aMC@NLO, which are then decayed with Pythia6 to all-fermion final states. Interference effects between the same fermion final states with distinct intermediate states is therefore not included, though MG5\_aMC@NLO is capable of doing this. Fermion pair, diboson and triboson ($2f,4f,6f$) states from initial $e^+e^-$ states are generated with MG5\_aMC@NLO, specifiying beam type 0 (no PDF) for both electron and positron. Inelastic Compton scattering processes ($1f,3f$) with final states $e\gamma,eZ,\nu W$ are generated by specifiying beam type 0 for the electron or positron (no PDF) and beam type 3 (photon PDF from electron beam) for the photon. In the latter case MG5\_aMC@NLO uses the effective photon approximation to simulate the Weizsacker-Williams photons generated by bremstrahlung. Finally, the $\gamma \gamma \rightarrow f\bar{f}, W^+ W^-$  ($aa2f,aa4f$) processes are simulated by specifiying beam type 3 for both electron and positron, using the effective photon approximation for both photons.

\begin{table*}[t]
\begin{center}
\begin{tabular}{|c|c|c|c|c|} \hline
$\sqrt{s}$[GeV] & Pol. & Process & $\sigma$[pb] CEPC & $\sigma$[pb] MG5 \\ \hline
250 & none & $e^+ e^- \rightarrow \mu^+ \mu^-, \tau^+ \tau^-$ & 4.40 & 3.50 \\
250 & none & $e^+ e^- \rightarrow q\bar{q}$ & 50.2 & 11.3 \\
250 & none & $e^+ e^- \rightarrow ZZ$ & 1.03 & 1.10 \\
250 & none & $e^+ e^- \rightarrow WW$ & 15.4 & 16.5 \\
250 & none & $e^+ e^- \rightarrow Zh$ & 0.212 & 0.240 \\ \hline \hline
$\sqrt{s}$[GeV] & Pol. & Process & $\sigma$[pb] ILC & $\sigma$[pb] MG5 \\ \hline
250 & ($+,-$) & $e^+ e^- \rightarrow Zh$ & 0.319 & 0.356 \\
250 & ($-,+$) & $e^+ e^- \rightarrow Zh$ & 0.206 & 0.240 \\
350 & ($+,-$) & $e^+ e^- \rightarrow t \bar{t}$ & 0.286 & 0.378 \\
350 & ($-,+$) & $e^+ e^- \rightarrow t \bar{t}$ & 0.137 & 0.166 \\
500 & ($+,-$) & $e^+ e^- \rightarrow t \bar{t}$ & 1.08 & 0.921 \\
500 & ($-,+$) & $e^+ e^- \rightarrow t \bar{t}$ & 0.470 & 0.436 \\ \hline
\end{tabular}
\caption{MG5\_aMC@NLO cross sections compared to ILC DBD \cite{Behnke:2013lya} and CEPC \cite{Mo:2015mza} cross sections. The former are generally larger than the latter due to beamstrahlung simulation in the latter. The discrepancy between the CEPC and MG5 cross sections for fermion pairs, which may be due to differing treatment of radiative return events, is under investigation.} 
\label{tab:compare}
\end{center}
\end{table*}

For each generated process with polarized beams, we generate a number of events whose equivalent luminosity is approximately five times the most optimistic operating scenario of scenarios G-20, H-20 and I-20, namely 10ab$^{-1}$ for each beam polarization at $\sqrt{s}=500$~GeV, 6.75ab$^{-1}$ for polarization $(+,-)$ at $\sqrt{s}=250,350$ and 2.25ab$^{-1}$ for polarization $(-,+)$ at $\sqrt{s}=250,350$. For each process with unpolarized beams at $\sqrt{s}=250$~GeV the luminosity is chosen to match the statistics of the polarized beam samples, and at $\sqrt{s}=91$~GeV the luminosity is chosen to be five times a GigaZ production, or $0.108$ab$^{-1}$.

In order to ensure that the process cross section converges, the event particles enter the effective detector radius, and to speed production, in some samples kinematic cuts have been applied to generator particles. In $Z\gamma$, $WW\gamma$ and $e\gamma$ processes, a photon requirement $p_{T}^{\gamma}>20$~GeV is imposed. In the $eZ$ and $\nu W$ samples requirements $p_{T}^{e}>20$~GeV and $p_{T}^{\nu}>20$~GeV are imposed. Finally, in the $Wee$ and $Zee$ samples a requirement $p_{T}^{e}>20$~GeV and $p_{T}^{e}>1$~keV, respectively, are imposed.

We perform fast detector simulation with both the MG5\_aMC@NLO samples and the DBD samples using Delphes3 and the DSiD detector card. The DSiD detector card is modeled on the full simulation performance of the SiD detector. The detector object efficiencies, fake rates and resolutions specified in the DSiD detector card can be reproduced in complex $e^+e^-$ event environments as demonstrated in the validation documentation \cite{Potter:2016pgp}.

\section{Background Analysis}

We emphasize that in this study both the DBD and MG5 samples are submitted to fast detector simulation with Delphes using the same DSiD card. Any difference between their distributions in the Delphes files cannot therefore be due to detector effects. The differences in generation between the DBD and CEPC samples and the MG5/DSiD samples described here, already discussed above, are here made explicit:

\begin{itemize}

\item beamstrahlung is not included in the MG5/DSiD samples; the beam energy distribution is idealized 

\item interference in distinct fermion final states between different intermediate bosonic states is included in the DBD and CEPC samples but is not in the MG5/DSiD samples

\item some MG5/DSiD samples include generator cuts on event particle $p_{T}$ to control divergent cross sections and ensure the particles enter the detector, while the DBD and CEPC samples include cuts on fermion pair invariant mass

\item all MG5/DSiD sample events are unweighted ($W=1$) so the samples should be scaled by cross section to the desired luminosity, while the DBD samples are weighted ($W\neq 1$) to achieve a target luminosity

\end{itemize}

\noindent Any background analysis which uses these samples should account for differences and assign any necessary uncertainties. 

\begin{figure*}[p]
\begin{center}
\includegraphics[width=0.4\textwidth]{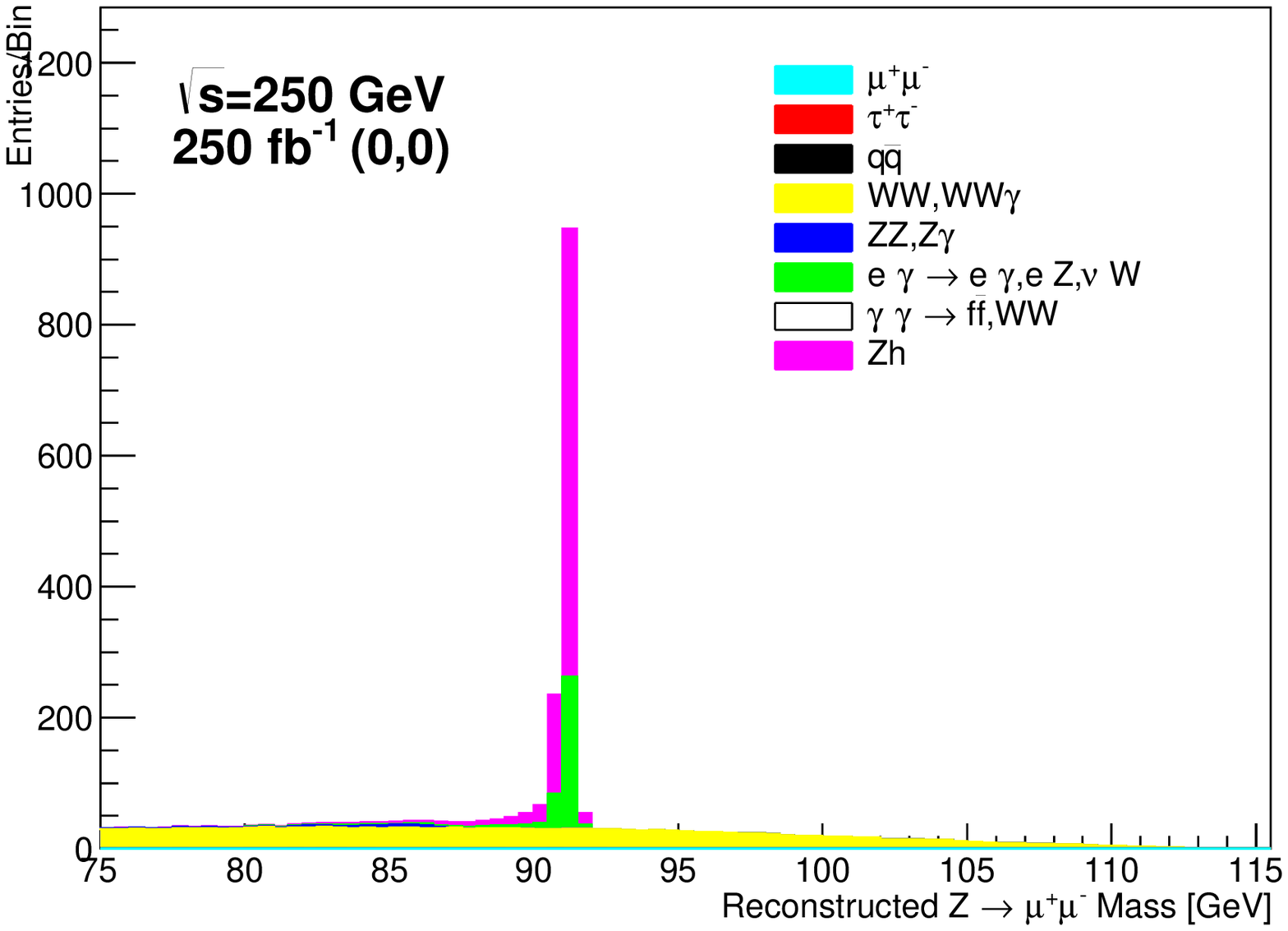}
\includegraphics[width=0.4\textwidth]{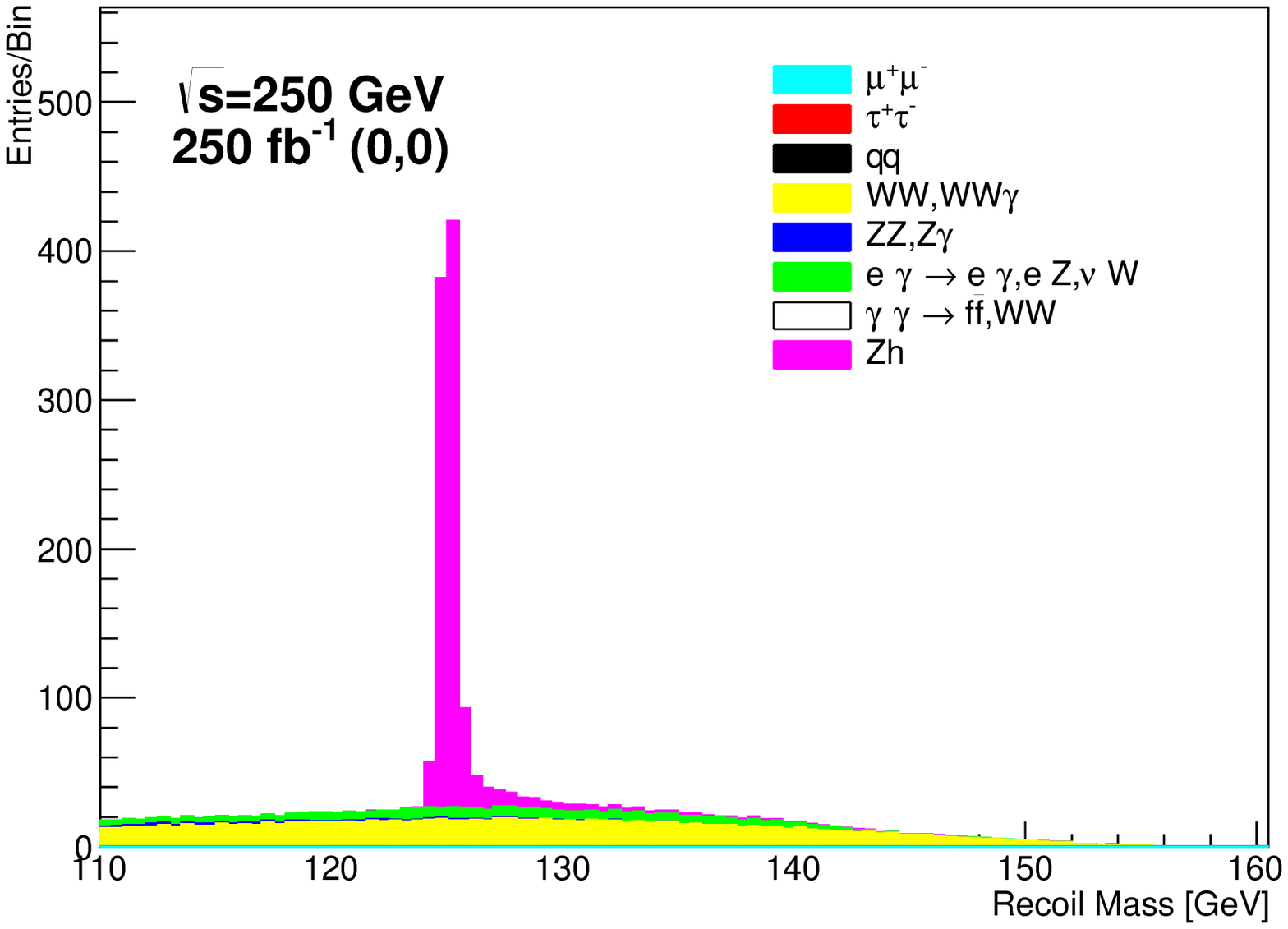}
\includegraphics[width=0.4\textwidth]{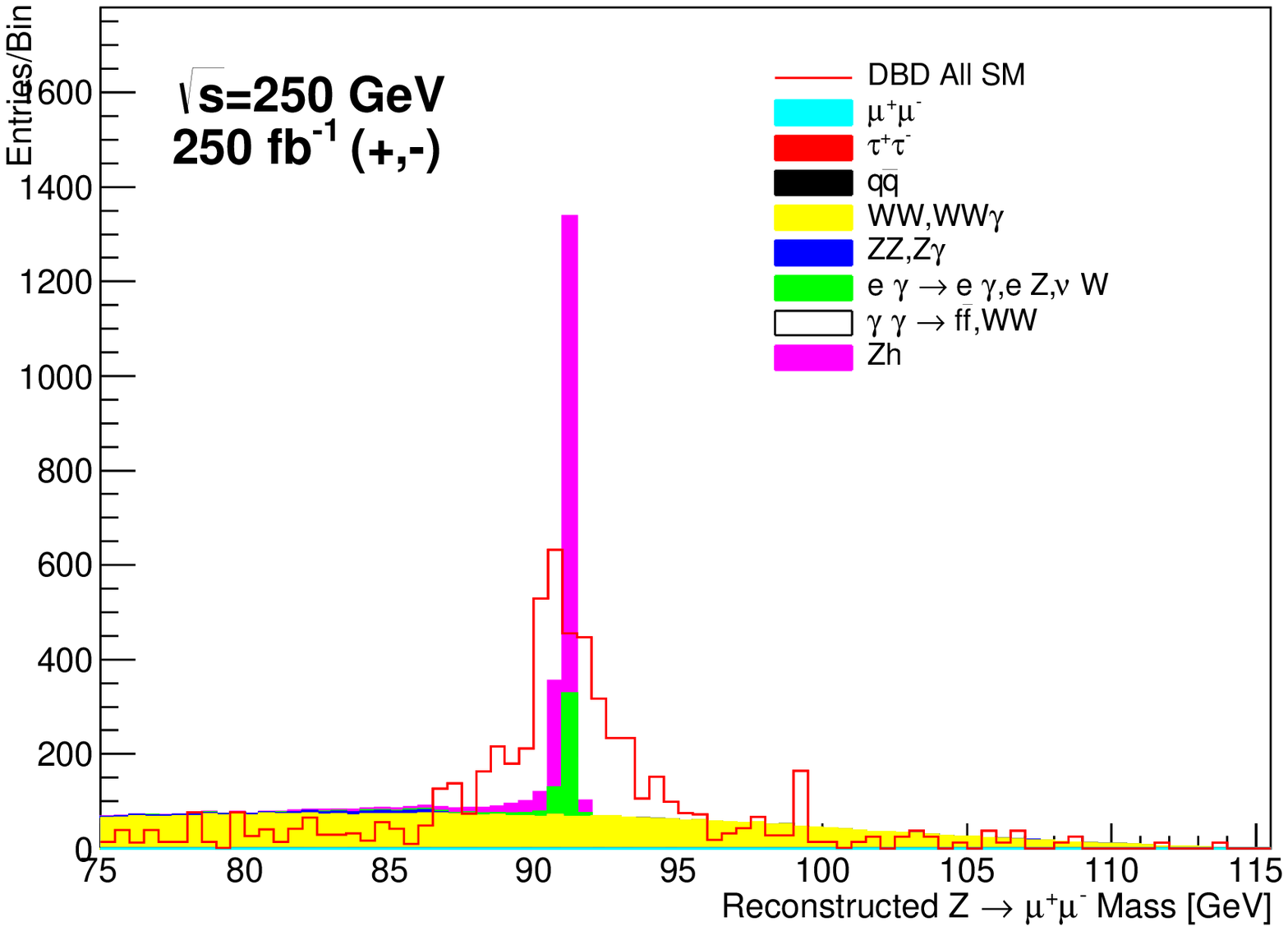}
\includegraphics[width=0.4\textwidth]{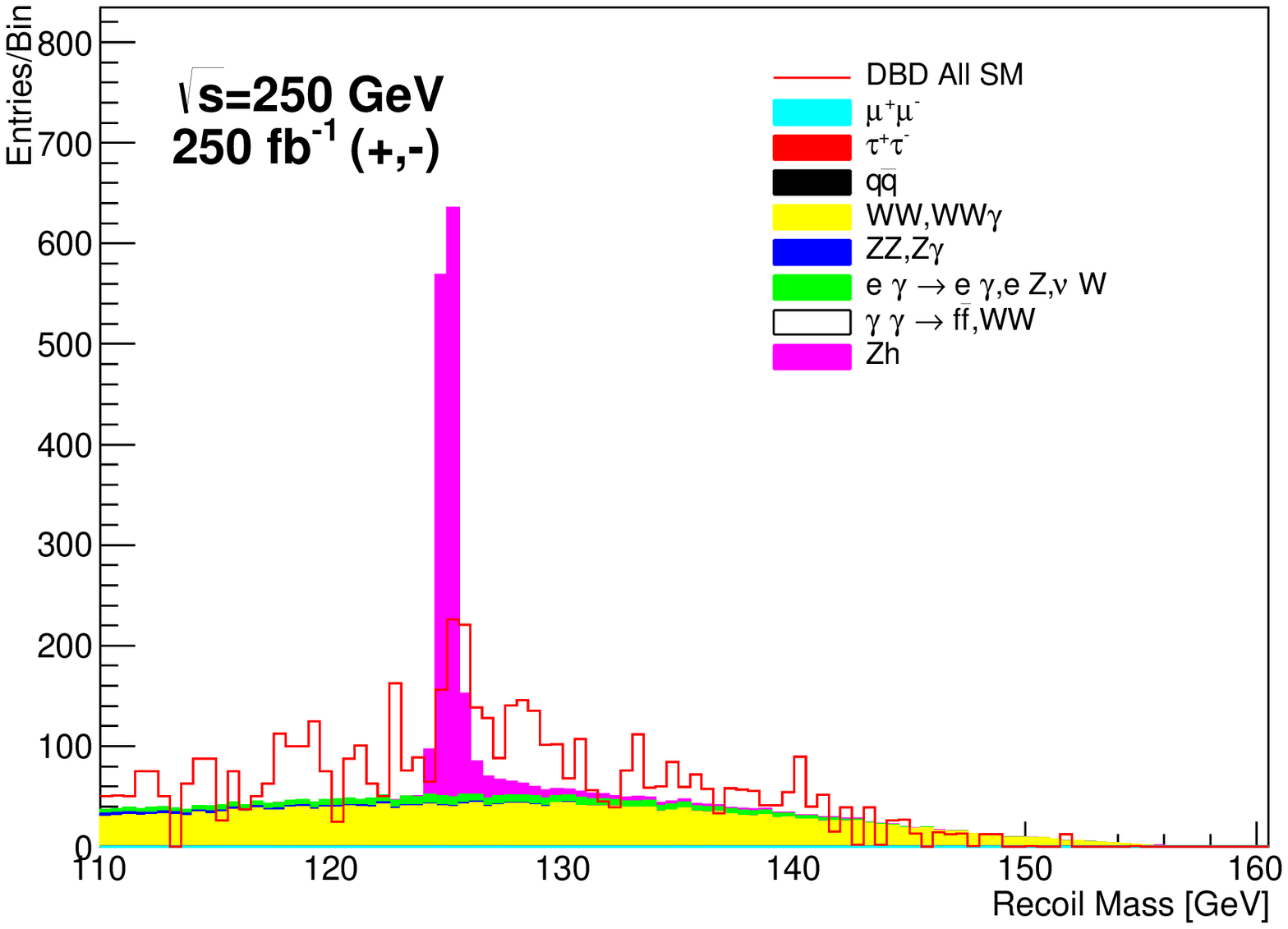}
\caption{Reconstructed $Z \rightarrow \mu^+ \mu^-$ and recoil masses for unpolarized beams at $\sqrt{s}=250$~GeV (above) and ($+,-$) beams at $\sqrt{s}=250$~GeV (below).  DBD distributions include \texttt{higgs\_ffh} and \texttt{all\_SM\_background}.}
\label{fig:reco1}
\end{center}
\begin{center}
\includegraphics[width=0.4\textwidth]{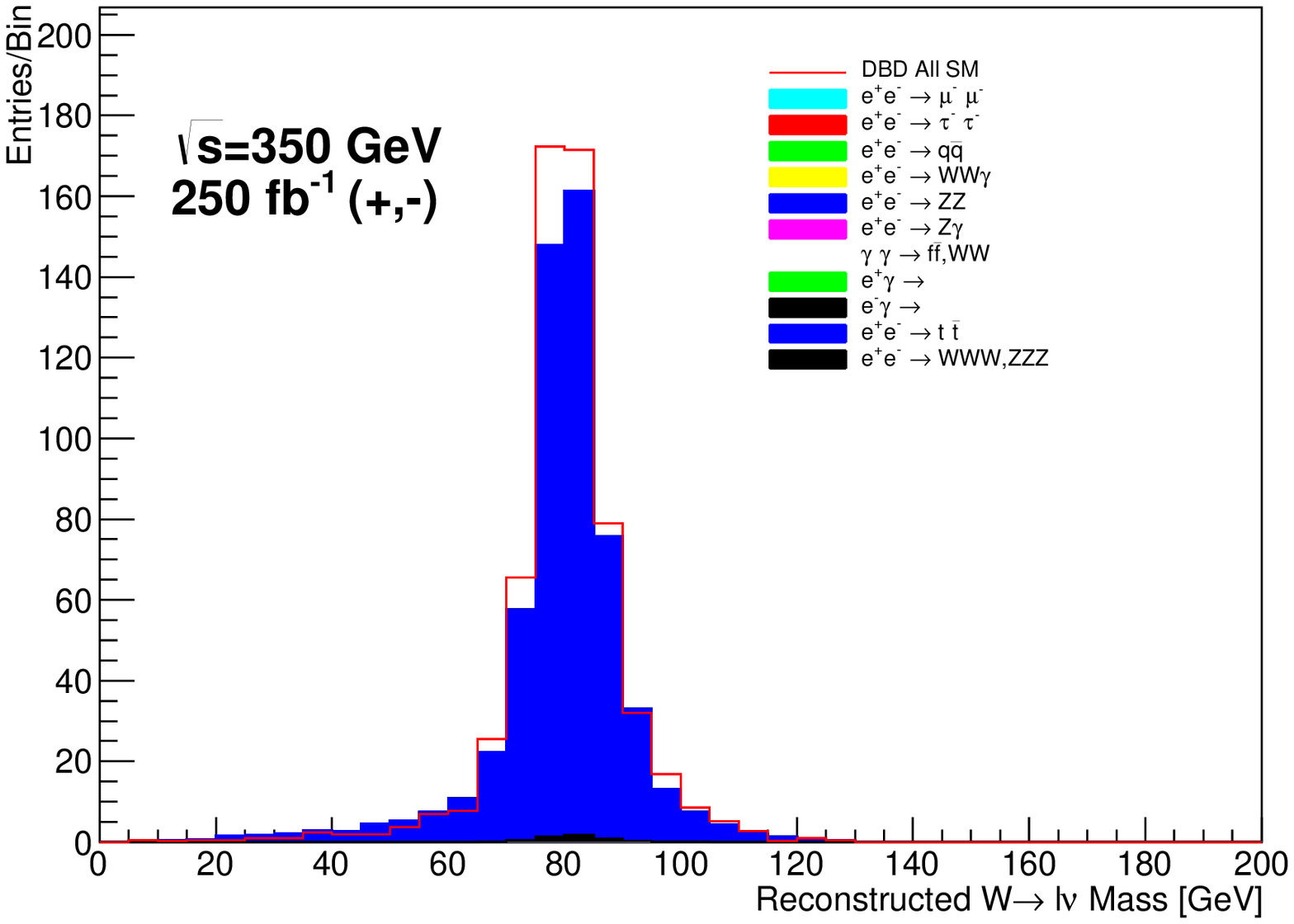}
\includegraphics[width=0.4\textwidth]{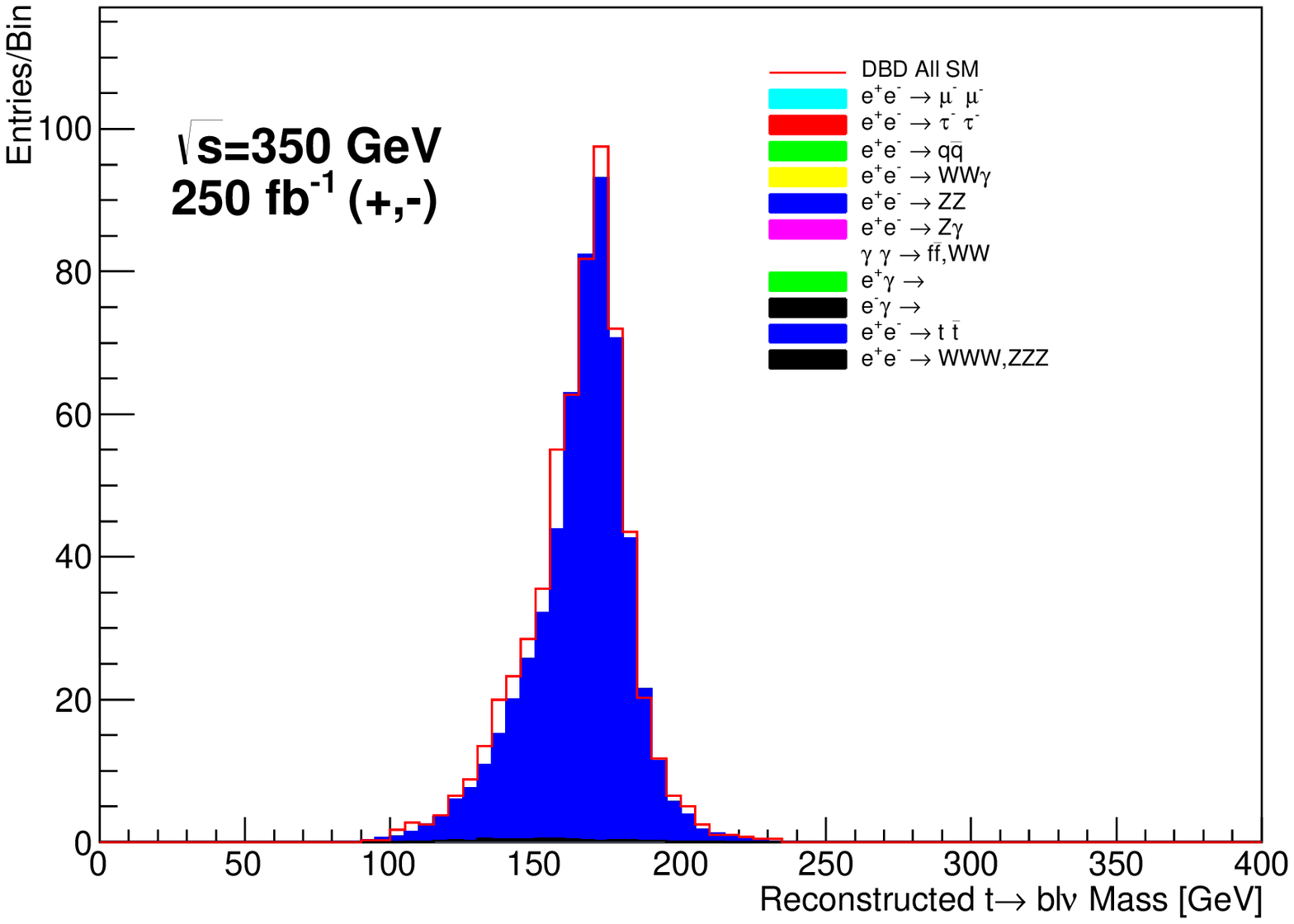}
\includegraphics[width=0.4\textwidth]{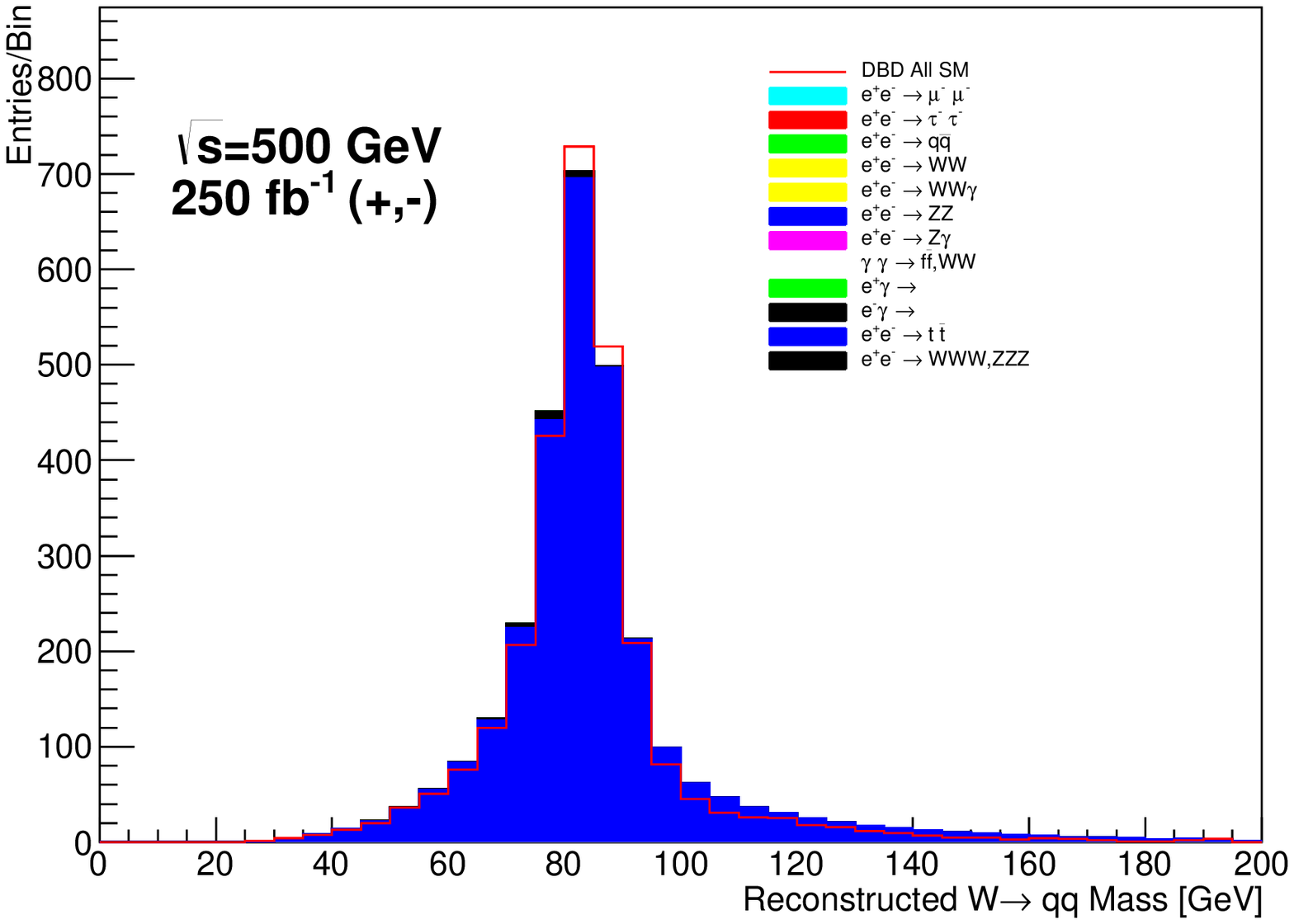}
\includegraphics[width=0.4\textwidth]{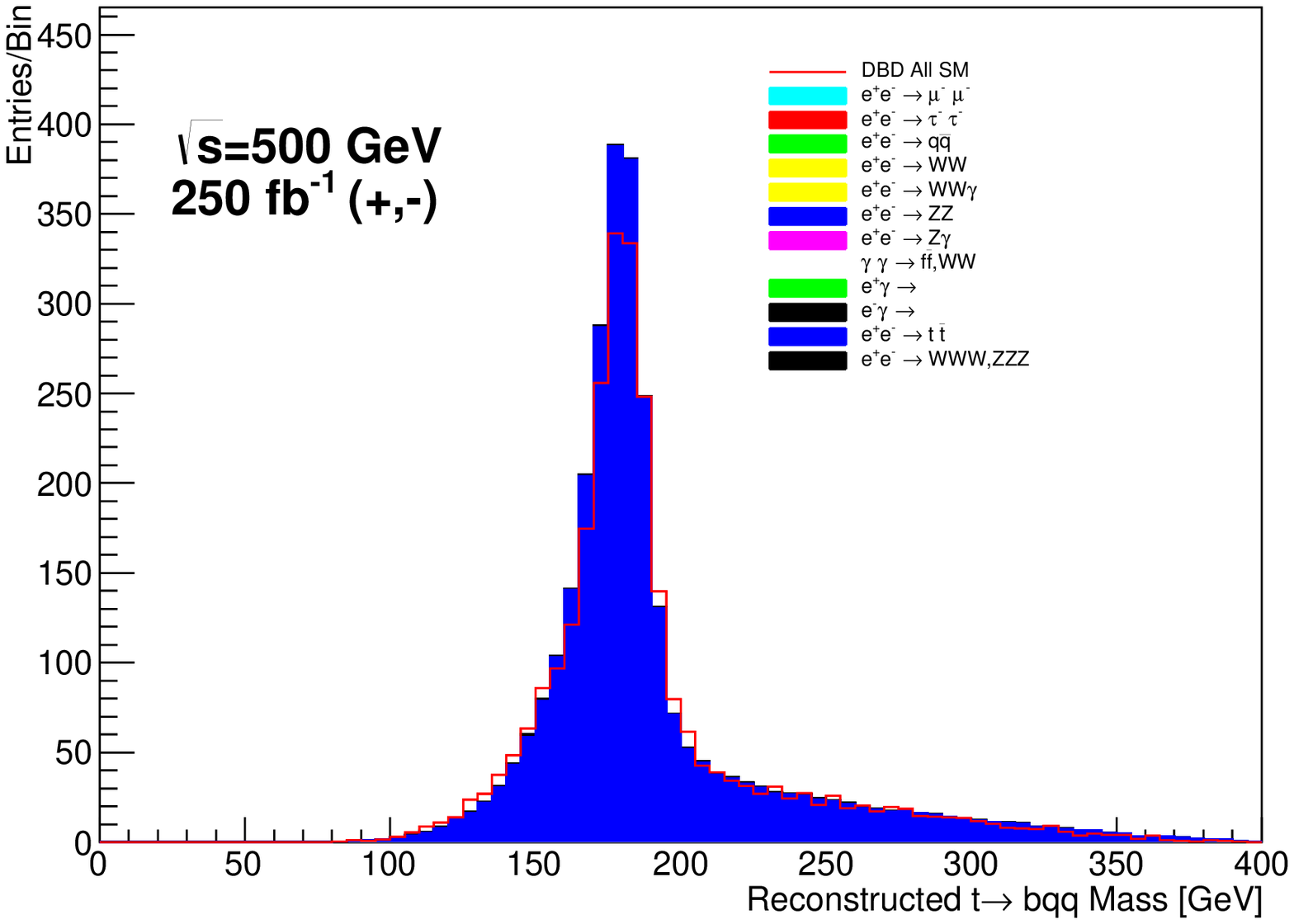}
\caption{Reconstructed $W_{l} \rightarrow \ell \nu$ and $t \rightarrow b W_{l}$ masses for ($+,-$) beams at $\sqrt{s}=350$~GeV (above), and $W_{h} \rightarrow q q^{\prime}$ and $t \rightarrow b W_{h} $ masses for ($+,-$) beams at $\sqrt{s}=500$~GeV (below). DBD distributions include \texttt{ttbar} and \texttt{all\_other\_SM\_background} (above), \texttt{6f\_ttbar\_mt173p5} and \texttt{all\_SM\_background} (below).}
\label{fig:reco2}
\end{center}
\end{figure*}

For illustration, we perform the following background studies. For each $\sqrt{s}$ we perform an analysis of the Delphes files with Root 5.34, exploiting multicore capability with Proof. At $\sqrt{s}=250$~GeV we  reconstruct the SM Higgs in $e^+ e^- \rightarrow Zh_{SM}$ using the recoil technique with $Z \rightarrow \mu^+ \mu^-$. The selection requires two oppositely charged, acollinear ($110^{\degree} < \theta < 150^{\degree}$) muons with $p_{T}^{\mu}>20$~GeV, $p_{T}^{\mu\mu}>20$~GeV and invariant mass $\vert m_{\mu^+ \mu^-}-m_{Z}\vert<10$~GeV. See Figure \ref{fig:reco1} for the $\sqrt{s}=250$~GeV reconstructed mass distributions.

For $\sqrt{s}=350$~GeV we reconstruct top pair events with $t \rightarrow bW$ in which one $W$ decays leptonically and the other $W$ decays hadronically. The signal selection requires exactly one lepton $\ell=e,\mu$  with $p_{T}^{\ell}>20$~GeV, missing transverse energy $E_{T}^{miss}>20$~GeV, at least four jets with $p_{T}^{j}>20$~GeV  exactly two of which must be $b$-tagged. The hadronic $W$ is reconstructed from the two leading untagged jets while the leptonic $W$ is reconstructed from the lepton and the missing energy. To reconstruct the top quarks, each $b$-jet is assigned to the reconstructed $W$ which maximizes the $\Delta R(b,W)$ since the top quarks are produced near threshold. For the $\sqrt{s}=500$~GeV sample we reconstruct top pair events exactly as for $\sqrt{s}=350$~GeV except that $b$-jets are assigned to the $W$ which \emph{minimizes} $\Delta R(b,W)$. See Figure \ref{fig:reco2} for the $\sqrt{s}=350,500$~GeV reconstructed mass distributions, where in both cases the distributions are normalized to the DBD cross sections in Table~\ref{tab:compare}.

Systematic and statistical uncertainties on the yields can be evaluated as follows.  Reduction in cross section due to beam energy loss can be estimated from Table~\ref{tab:compare} to be 10\% and 25\%, respectively, for the Higgs recoil and top pair analyses. Moreover, the recoil mass is smeared by the beam energy distribution if radiative losses are not recovered. Interference between intermediate states affects both the Higgs recoil (intermediate $WW$ and $ZZ$ states with $\mu^+ \mu^- \nu_{\mu} \bar{\nu}_{\mu}$ final state) and top pair analyses (intermediate $WWZ$ and $t\bar{t}$ states with $W^+W^-b\bar{b}$ states). For the Higgs recoil analysis, the cross sections for polarization ($+,-$) are calculated by MG5\_aMC@NLO to b 8.4fb$^{-1}$ for $ZZ\rightarrow \mu^+ \mu^- \nu_{\mu} \bar{\nu}_{\mu} $, 453.9fb$^{-1}$ for $WW \rightarrow \mu^+ \mu^- \nu_{\mu} \bar{\nu}_{\mu}$ and 462.4fb$^{-1}$ for $\mu^+ \mu^- \nu_{\mu} \bar{\nu}_{\mu}$ including interference. The relative uncertainty is therefore below 1\%, the reported generator uncertainty. For the top pair analysis the effect is negligible compared to the uncertainty introduced by beam energy distribution.
Since background from $WW\gamma$ and $Z\gamma$ events with $p_{T}^{\gamma}<20$~GeV, as well as $eZ$ events with $p_{T}^{e}<20$~GeV,  avoid the generator level requirements at $\sqrt{s}=250$~GeV, some background is neglected.  Because $WWZ$ events at $\sqrt{s}=350,500$~GeV are generated without kinematic constraints, any omitted top pair background is negligible. The statistical uncertainty on the yields are computed straightforwardly and scale as $N^{-1/2}$. Comparing with a DBD sample with integrated luminosity $0.25$ab$^{-1}$ and event weight $W=1$, the improvement in statistical uncertainty is approximately $ \times 4.5$. For events with large weights $W>1$, the improvement factor is approximately $\sqrt{20W}$.

\section{Conclusion}

We have described the production of fast simulation background samples for new physics studies at a future $e^+ e^-$ collider like the ILC or CEPC. Events are generated for a variety of run scenarios with approximately five times the integrated luminosity envisaged by the most optimistic run scenario for each $\sqrt{s}$. The events are generated with MG5\_aMC@NLO with detector simulation performed by Delphes using the DSiD detector card. Finally, the samples are compared to the ILC background samples made for the DBD study and CEPC background samples.

Systematic uncertainties associated with the MG5\_aMC@NLO samples have been estimated. These samples lack a detailed simulation of initial state radiation and beamstrahlung. The $2f$ background from radiative return events is absent, and both pileup from bunch-bunch interactions and a realistic beam energy distribution are absent. Nevertheless, these shortcomings can be ameliorated. Moreover, the MG5/DSiD samples compare favorably to the DBD and CEPC in statistical uncertainty due to the large integrated luminosities and unweighted events in the MG5/DSiD samples.

\begin{center}\textbf{Acknowledgements}\end{center} The author thanks Jan Strube and Tomohiko Tanabe for feedback on an early draft of this paper, the Alder Institute for High Energy Physics for financial support, and the HKUST Institute for Advanced Study Program on High Energy Physics 2017 for travel support.

\bibliography{paper}

\begin{table*}[t]
\begin{center}
\begin{tabular}{|l|c|c|c|c|c|c|} \hline
Sample & Final State & Pol. & Cuts & $\sigma$ [pb] & $N$[M] & $\int \mathcal{L}$ [ab$^{-1}$] \\ \hline \hline
\texttt{ff91ub} & $f\bar{f}$ & none & none & $4.63 \times 10^4$  & 5000 & 0.108 \\
\texttt{ff91pm} & $f\bar{f}$ & $(+,-)$ & none & $6.86 \times 10^4$  & 2500 & 0.0364 \\
\texttt{ff91mp} & $f\bar{f}$ & $(-,+)$ & none & $4.63 \times 10^4$  & 2500 & 0.0540 \\ \hline
\texttt{aeffe91ub} & $e^- \gamma,e^- f\bar{f}$ & none & $p_{T}^{\ell}>20$~GeV  & $4.34$  & 1  & 0.230 \\
\texttt{aeffe91pm} & $e^- \gamma,e^- f\bar{f}$ & $(+,-)$ &  $p_{T}^{\ell}>20$~GeV & $4.34$  & 1  & 0.230 \\
\texttt{aeffe91mp} & $e^- \gamma,e^- f\bar{f}$ & $(-,+)$ &  $p_{T}^{\ell}>20$~GeV & $4.34$  & 1  & 0.230 \\ \hline
\texttt{papff91ub} & $e^+ \gamma, e^+ f\bar{f}$ & none &  $p_{T}^{\ell}>20$~GeV & $4.34$ & 1  &  0.230 \\
\texttt{papff91pm} & $e^+ \gamma, e^+ f\bar{f}$ & $(+,-)$ &  $p_{T}^{\ell}>20$~GeV & $4.34$ & 1  &  0.230 \\
\texttt{papff91mp} & $e^+ \gamma, e^+ f\bar{f}$ & $(-,+)$ &  $p_{T}^{\ell}>20$~GeV & $4.34$ & 1  &  0.230 \\ \hline
\texttt{aaff91ub} & $\gamma \gamma \rightarrow f\bar{f}$ & none & none  & $5.92 \times 10^2$ & 64 & 0.108\\ 
\texttt{aaff91pm} & $\gamma \gamma \rightarrow f\bar{f}$ & $(+,-)$ & none  & $5.92 \times 10^2$ & 32& 0.054\\ 
\texttt{aaff91mp} & $\gamma \gamma \rightarrow f\bar{f}$ & $(-,+)$ & none  & $5.92 \times 10^2$ & 32& 0.054 \\ \hline
\end{tabular}
\caption{For the $\sqrt{s}=91$~GeV samples: processes, polarization, generator cuts, MG5\_aMC@NLO cross section, number of events generated and equivalent integrated luminosity. Here $f=\mu\tau udscb$.} 
\label{tab:91}
\end{center}
\end{table*}

\begin{table*}[p]
\begin{center}
\begin{tabular}{|l|c|c|c|c|c|c|} \hline
Sample & Final State & Pol. & Cuts & $\sigma$ [pb] & $N$[M] & $\int \mathcal{L}$ [ab$^{-1}$] \\ \hline \hline

\texttt{mumu250ub} & $ \mu^+ \mu^-$ & none & none & 1.73 & 21 & 12.1 \\
\texttt{mumu250pm} & $ \mu^+ \mu^-$ & ($+,-$) & none &  2.36 & 16 & 6.78 \\
\texttt{mumu250mp} & $ \mu^+ \mu^-$ & ($-,+$) & none &  1.94 & 5  & 2.58 \\ \hline
\texttt{tautau250ub} & $ \tau^+ \tau^-$ & none & none & 1.77 &  21  &  11.9\\
\texttt{tautau250pm} & $ \tau^+ \tau^-$ & ($+,-$) & none &  2.41 & 17 & 7.05 \\
\texttt{tautau250mp} & $ \tau^+ \tau^-$ & ($-,+$) & none &  1.98 & 5  & 2.53 \\ \hline
\texttt{qq250ub} & $ q\bar{q}$ $(q=udscb)$ & none & none &  11.3 & 156  & 13.8 \\
\texttt{qq250pm} & $ q\bar{q}$ $(q=udscb)$ & ($+,-$) & none &  20.4 & 138 & 6.76 \\
\texttt{qq250mp} & $ q\bar{q}$ $(q=udscb)$ & ($-,+$) & none &  7.64 &  18 & 2.36 \\ \hline
\texttt{zz250ub} & $ ZZ$ & none & none & 1.10 & 15& 13.6\\
\texttt{zz250pm} & $ ZZ$ & ($+,-$) & none &  1.87 & 13 & 6.95 \\
\texttt{zz250mp} & $ ZZ$ & ($-,+$) & none &  0.858 &  2 & 2.33 \\ \hline
\texttt{zh250ub} & $ Zh$ & none & none & 0.240 &  4& 16.7 \\
\texttt{zh250pm} & $ Zh$ & ($+,-$) & none & 0.356  & 3 & 8.43 \\
\texttt{zh250mp} & $ Zh$ & ($-,+$) & none & 0.240  & 1  & 4.17  \\ \hline
\texttt{za250ub} & $ Z\gamma$ & none & $p_{T}^{\gamma}>20$~GeV & 7.71 &  95&  12.3\\
\texttt{za250pm} & $ Z\gamma$ & ($+,-$) & $p_{T}^{\gamma}>20$~GeV &  11.4 & 77&  6.75\\
\texttt{za250mp} & $ Z\gamma$ & ($-,+$) & $p_{T}^{\gamma}>20$~GeV &  7.70 &  18 & 2.34 \\ \hline
\texttt{ww250ub} & $ W^+ W^-$ & none & none &  16.5 & 266  & 16.1\\
\texttt{ww250pm} & $ W^+ W^-$ & ($+,-$) & none &  38.3 & 259 & 6.76 \\
\texttt{ww250mp} & $ W^+ W^-$ & ($-,+$) & none &  2.63 &  6 &  2.28\\ \hline
\texttt{wwa250ub} & $ W^+ W^- \gamma$ & none & $p_{T}^{\gamma}>20$~GeV & 0.121 &  3  & 24.8 \\
\texttt{wwa250pm} & $ W^+ W^- \gamma$ & ($+,-$) & $p_{T}^{\gamma}>20$~GeV &  0.278 & 2 & 7.19\\
\texttt{wwa250mp} & $ W^+ W^- \gamma$ & ($-,+$) & $p_{T}^{\gamma}>20$~GeV &  0.021 & 1  & 47.6 \\ \hline
\texttt{zeezvv250ub} & $ Zee,Z\nu_e \nu_e$ & none & $p_{T}^{e}>20$~GeV & 0.332 & 5 & 15.1 \\
\texttt{zeezvv250pm} & $ Zee,Z\nu_e \nu_e$ & ($+,-$) & $p_{T}^{e}>20$~GeV & 0.574  &  4 &  6.97\\
\texttt{zeezvv250mp} & $ Zee,Z\nu_e\nu_e$ & ($-,+$) & $p_{T}^{e}>20$~GeV & 0.233 & 1  & 4.29 \\ \hline 
\texttt{wev250ub} & $ We\nu_e$ & none & $p_{T}^{e}>20$~GeV & 3.53  & 57 & 16.1 \\
\texttt{wev250pm} & $ We\nu_e$ & ($+,-$) & $p_{T}^{e}>20$~GeV & 8.15  & 55 &  6.75\\
\texttt{wev250mp} & $ We\nu_e$ & ($-,+$) & $p_{T}^{e}>20$~GeV & 0.579  & 2  & 3.45 \\ \hline \hline
\texttt{papzwv250ub} & $e^+ \gamma, e^+ Z,\nu W^+$ & none & $p_{T}^{\gamma,\nu,e}>20$~GeV  & 10.8 & 99 & 9.17  \\
\texttt{papzwv250pm} & $e^+ \gamma, e^+ Z,\nu W^+$ & ($+,-$) & $p_{T}^{\gamma,\nu,e}>20$~GeV  & 11.0 & 75 & 6.82 \\
\texttt{papzwv250mp} & $e^+ \gamma, e^+ Z,\nu W^+$  & ($-,+$) & $p_{T}^{\gamma,\nu,e}>20$~GeV  & 10.6 & 24&  2.26\\ \hline
\texttt{aezewv250ub} & $e^- \gamma,e^- Z,\nu W^-$ & none & $p_{T}^{\gamma,\nu,e}>20$~GeV  & 10.8 & 100 & 9.26 \\
\texttt{aezewv250pm} & $e^- \gamma,e^- Z,\nu W^-$ & ($+,-$) & $p_{T}^{\gamma,\nu,e}>20$~GeV  & 11.3 & 77 & 6.81 \\
\texttt{aezewv250mp} & $e^- \gamma, e^- Z,\nu W^-$  & ($-,+$) & $p_{T}^{\gamma,\nu,e}>20$~GeV  & 10.2 &  23&  2.25\\ \hline \hline
\texttt{aaffww250ub} & $\gamma \gamma \rightarrow f\bar{f},WW$ & none & $p_{T}^{f}>20$~GeV  & 2.27 &  22&  9.69\\
\texttt{aaffww250pm} & $\gamma \gamma \rightarrow f\bar{f},WW$ & ($+,-$) & $p_{T}^{f}>20$~GeV  & 2.27 &  16& 7.05 \\
\texttt{aaffww250mp} & $\gamma \gamma \rightarrow f\bar{f},WW$  & ($-,+$) & $p_{T}^{f}>20$~GeV  & 2.27 & 6 & 2.64\\ \hline

\end{tabular}
\caption{For the $\sqrt{s}=250$~GeV samples: processes, polarization, generator cuts, MG5\_aMC@NLO cross section, number of events generated and equivalent integrated luminosity.} 
\label{tab:250}
\end{center}
\end{table*}

\begin{table*}[p]
\begin{center}
\begin{tabular}{|l|c|c|c|c|c|c|} \hline
Sample & Final State & Pol. & Cuts & $\sigma$ [pb] & $N$[M] & $\int \mathcal{L}$ [ab$^{-1}$] \\ \hline \hline

\texttt{mumu350pm} & $ \mu^+ \mu^-$ & ($+,-$) & none &  1.19& 8& 6.72 \\
\texttt{mumu350mp} & $ \mu^+ \mu^-$ & ($-,+$) & none &  0.992& 3 &3.02  \\ \hline
\texttt{tautau350pm} & $ \tau^+ \tau^-$ & ($+,-$) & none & 1.19 &8& 6.72\\
\texttt{tautau350mp} & $ \tau^+ \tau^-$ & ($-,+$) & none & 0.994 &3 &3.02\\ \hline
\texttt{qq350pm} & $ q\bar{q}$ $(q=udscb)$ & ($+,-$) & none & 9.67 &66& 6.83\\
\texttt{qq350mp} & $ q\bar{q}$ $(q=udscb)$ & ($-,+$) & none & 3.69 &9&2.44\\ \hline
\texttt{tt350pm} & $ t\bar{t}$ & ($+,-$) & none &  0.378 & 3& 7.94  \\
\texttt{tt350mp} & $ t\bar{t}$ & ($-,+$) & none & 0.166 & 1& 6.02 \\ \hline
\texttt{zz350pm} & $ ZZ$ & ($+,-$) & none & 1.16  & 8 & 6.90\\
\texttt{zz350mp} & $ ZZ$ & ($-,+$) & none & 0.532 & 2 & 3.76\\ \hline
\texttt{za350pm} & $ Z\gamma$ & ($+,-$) & $p_{T}^{\gamma}>20$~GeV & 6.27 & 43& 6.86 \\
\texttt{za350mp} & $ Z\gamma$ & ($-,+$) & $p_{T}^{\gamma}>20$~GeV & 4.23 & 10 & 2.36\\ \hline
\texttt{ww350pm} & $ W^+ W^-$ & ($+,-$) & none & 26.3 & 178 & 6.77\\
\texttt{ww350mp} & $ W^+ W^-$ & ($-,+$) & none & 1.73 & 4& 2.31\\ \hline
\texttt{wwa350pm} & $ W^+ W^- \gamma$ & ($+,-$) & $p_{T}^{\gamma}>20$~GeV & 0.397 & 3 & 7.56 \\
\texttt{wwa350mp} & $ W^+ W^- \gamma$ & ($-,+$) & $p_{T}^{\gamma}>20$~GeV & 0.030 & 1 & 33.3\\ \hline
\texttt{zeezvv350pm} & $ Zee,Z\nu_e \nu_e$ & ($+,-$) & $p_{T}^{e}>20$~GeV &  0.702 & 5 & 7.12 \\
\texttt{zeezvv350mp} & $ Zee,Z \nu_e \nu_e$ & ($-,+$) & $p_{T}^{e}>20$~GeV & 0.222  & 1  & 4.50 \\ \hline
\texttt{wev350pm} & $ We\nu_e$ & ($+,-$) & $p_{T}^{e}>20$~GeV & 6.47  & 44 &  6.80\\
\texttt{wev350mp} & $ We\nu_e$ & ($-,+$) & $p_{T}^{e}>20$~GeV & 0.486  & 1  & 2.06 \\ \hline
\texttt{vvv350pm} & $ WWZ,ZZZ$ & ($+,-$) & none  &  0.030 & 1 &  33.3 \\
\texttt{vvv350mp} & $ WWZ,ZZZ$ & ($-,+$) & none  &  0.003 & 1 &  333. \\ \hline \hline
\texttt{papzwv350pm} & $e^+ \gamma, e^+ Z,\nu W^+$ & ($+,-$) & $p_{T}^{\gamma,\nu,e}>20$~GeV  & 13.4 & 91 & 6.79 \\
\texttt{papzwv350mp} & $e^+ \gamma, e^+ Z,\nu W^+$  & ($-,+$) & $p_{T}^{\gamma,\nu,e}>20$~GeV  & 12.6 & 29 & 2.30 \\ \hline
\texttt{aezewv350pm} & $e^- \gamma, e^- Z,\nu W^-$ & ($+,-$) & $p_{T}^{\gamma,\nu,e}>20$~GeV  & 14.2  & 96 & 6.76 \\
\texttt{aezewv350mp} & $e^- \gamma, e^- Z,\nu W^-$  & ($-,+$) & $p_{T}^{\gamma,\nu,e}>20$~GeV  & 11.8 & 27& 2.29\\ \hline \hline
\texttt{aaffww350pm} & $\gamma \gamma \rightarrow f\bar{f},WW$ & ($+,-$) & $p_{T}^{f}>20$~GeV  & 3.52 & 24 & 6.82 \\
\texttt{aaffww350mp} & $\gamma \gamma \rightarrow f\bar{f},WW$  & ($-,+$) & $p_{T}^{f}>20$~GeV  & 3.52 & 8 & 2.27 \\ \hline
\end{tabular}
\caption{For the $\sqrt{s}=350$~GeV samples: processes, polarization, generator cuts, MG5\_aMC@NLO cross section, number of events generated and equivalent integrated luminosity.} 
\label{tab:350}
\end{center}
\end{table*}

\begin{table*}[p]
\begin{center}
\begin{tabular}{|l|c|c|c|c|c|c|} \hline
Sample & Final State & Pol. & Cuts & $\sigma$ [pb] & $N$[M] & $\int \mathcal{L}$ [ab$^{-1}$] \\ \hline \hline

\texttt{mumu500pm} & $ \mu^+ \mu^-$ & ($+,-$) & none &  0.575 & 6 & 10.4  \\
\texttt{mumu500mp} & $ \mu^+ \mu^-$ & ($-,+$) & none &  0.482 &  5 & 10.4 \\ \hline
\texttt{tautau500pm} & $ \tau^+ \tau^-$ & ($+,-$) & none &  0.578 & 6& 10.4 \\
\texttt{tautau500mp} & $ \tau^+ \tau^-$ & ($-,+$) & none &  0.484 & 5  & 10.4 \\ \hline
\texttt{qq500pm} & $ q\bar{q}$ $(q=udscb)$ & ($+,-$) & none &  4.56 & 46 & 10.1 \\
\texttt{qq500mp} & $ q\bar{q}$ $(q=udscb)$ & ($-,+$) & none &  1.76 & 18 & 10.2 \\ \hline
\texttt{tt500pm} & $ t\bar{t}$ & ($+,-$) & none &  0.921 & 9 & 9.77 \\
\texttt{tt500mp} & $ t\bar{t}$ & ($-,+$) & none &  0.436 & 5  & 11.5 \\ \hline
\texttt{zz500pm} & $ ZZ$ & ($+,-$) & none &  0.707 &  7&  9.90\\
\texttt{zz500mp} & $ ZZ$ & ($-,+$) & none &  0.324 &  3& 9.26 \\ \hline
\texttt{za500pm} & $ Z\gamma$ & ($+,-$) & $p_{T}^{\gamma}>20$~GeV &  2.48 & 25 & 10.1 \\
\texttt{za500mp} & $ Z\gamma$ & ($-,+$) & $p_{T}^{\gamma}>20$~GeV &  1.67 & 17 & 10.2\\ \hline
\texttt{ww500pm} & $ W^+ W^-$ & ($+,-$) & none &  16.8 & 197& 11.7 \\
\texttt{ww500mp} & $ W^+ W^-$ & ($-,+$) & none &  1.07 &  11 & 10.2 \\ \hline
\texttt{wwa500pm} & $ W^+ W^- \gamma$ & ($+,-$) & $p_{T}^{\gamma}>20$~GeV & 0.360 & 4 & 11.1 \\
\texttt{wwa500mp} & $ W^+ W^- \gamma$ & ($-,+$) & $p_{T}^{\gamma}>20$~GeV & 0.026 & 1& 38.5\\ \hline
\texttt{zeezvv500pm} & $ Zee,Z\nu_e\nu_e$ & ($+,-$) & $p_{T}^{e}>20$~GeV & 1.06  & 11 & 10.4 \\
\texttt{zeezvv500mp} & $ Zee,Z\nu_e\nu_e$ & ($-,+$) & $p_{T}^{e}>20$~GeV &  0.235 &  3 &  12.8\\  \hline
\texttt{wev500pm} & $We\nu_e$ & ($+,-$) & $p_{T}^{e}>20$~GeV &  5.51 & 55 & 9.98 \\
\texttt{wev500mp} & $We\nu_e$ & ($-,+$) & $p_{T}^{e}>20$~GeV & 0.481  & 5  &  10.4\\ \hline
\texttt{vvv500pm} & $ WWZ,ZZZ$ & ($+,-$) &  none & 0.094 & 1 & 10.6 \\
\texttt{vvv500mp} & $ WWZ,ZZZ$ & ($-,+$) &  none & 0.007 & 1 & 143 \\ \hline \hline
\texttt{papzwv500pm} & $e^+ \gamma, e^+ Z,\nu W^+$ & ($+,-$) & $p_{T}^{\gamma,\nu,e}>20$~GeV  &  16.3 & 163 & 10.0 \\
\texttt{papzwv500mp} & $e^+ \gamma, e^+ Z,\nu W^+$  & ($-,+$) & $p_{T}^{\gamma,\nu,e}>20$~GeV  &  14.7 & 147 & 10.0 \\ \hline
\texttt{aezewv500pm} & $e^- \gamma, e^- Z,\nu W^-$ & ($+,-$) & $p_{T}^{\gamma,\nu,e}>20$~GeV  & 17.6 & 176 & 10.0\\
\texttt{aezewv500mp} & $e^- \gamma, e^- Z,\nu W^-$  & ($-,+$) & $p_{T}^{\gamma,\nu,e}>20$~GeV  & 13.3 & 135 & 10.2\\ \hline \hline
\texttt{aaffww500pm} & $\gamma \gamma \rightarrow f\bar{f},WW$ & ($+,-$) & $p_{T}^{q}>20$~GeV  & 5.21 & 53 & 10.2 \\
\texttt{aaffww500mp} & $\gamma \gamma \rightarrow f\bar{f},WW$  & ($-,+$) & $p_{T}^{q}>20$~GeV  & 5.21  & 53 & 10.2\\ \hline

\end{tabular}
\caption{For the $\sqrt{s}=500$~GeV samples: processes, polarization, generator cuts, MG5\_aMC@NLO cross section, number of events generated and equivalent integrated luminosity.} 
\label{tab:500}
\end{center}
\end{table*}

\end{document}